\documentclass[pdflatex,sn-basic]{sn-jnl}

\usepackage{graphicx}%
\usepackage{tabularx}
\usepackage{multirow}%
\usepackage{amsmath,amssymb,amsfonts}%
\usepackage{amsthm}%
\usepackage{mathrsfs}%
\usepackage[title]{appendix}%
\usepackage{xcolor}%
\usepackage{textcomp}%
\usepackage{manyfoot}%
\usepackage{booktabs}%
\usepackage{algorithm}%
\usepackage{algorithmicx}%
\usepackage{algpseudocode}%
\usepackage{listings}%
\usepackage{makecell}
\usepackage{subcaption}
\usepackage{changepage}

\theoremstyle{thmstyleone}%
\theoremstyle{thmstyletwo}%

\theoremstyle{thmstylethree}%

\begin{document}

\title[Similarities in the Initiation of Upward Positive and Downward Negative Lightning Flashes]{Similarities in the Initiation of Upward Positive and Downward Negative Lightning Flashes} 


\author*[1]{\fnm{Toma} \sur{Oregel-Chaumont}}\email{toma.chaumont@epfl.ch}

\author[1]{\fnm{Mohammad} \sur{Azadifar}}\email{mohammad.azadifar@epfl.ch}

\author[2]{\fnm{Antonio} \sur{\v{S}unjerga}}\email{asunje00@fesb.hr}

\author[3]{\fnm{Marcos} \sur{Rubinstein}}\email{marcos.rubinstein@heig-vd.ch}

\author[1]{\fnm{Farhad} \sur{Rachidi}}\email{farhad.rachidi@epfl.ch}

\affil*[1]{\orgdiv{Electromagnetic Compatibility Laboratory}, \orgname{Swiss Federal Institute of Technology (EPFL)}, \orgaddress{\street{} \city{Lausanne}, \postcode{1015}, \state{VD}, \country{Switzerland}}}

\affil[2]{\orgdiv{Faculty of Electrical Engineering}, \orgname{University of Split}, \orgaddress{\street{} \city{Split}, \postcode{21000}, \state{}, \country{Croatia}}}

\affil[3]{\orgdiv{HEIG}, \orgname{University of Applied Sciences and Arts Western Switzerland}, \orgaddress{\street{} \city{Yverdon-les-Bains}, \postcode{1401}, \state{VD}, \country{Switzerland}}}



\abstract{This study examines the relationship between upward negative stepped leader pulses in upward positive lightning and preliminary breakdown pulses (PBPs) in downward negative lightning discharges.
Through analysis of simultaneous channel-base current and electric field data from the S\"antis tower in Switzerland, we found notable similarities between the ``Category A'' and ``B'' pulses associated with the initial continuous current of upward negative leaders, and the ``Classical'' and ``Narrow'' PBPs observed in downward negative flashes. 
Statistical comparisons reveal correlations between measured electric field and tower current parameters for Category A pulses, supporting the field--current relationship for preliminary breakdown proposed in previous studies. 
These results suggest 
that similar physical processes may be involved in the early stages of negative leader development in both upward and downward lightning, providing valuable insights into lightning initiation that could not be obtained from conventional field measurements alone.
Furthermore,  high-speed camera footage revealed that Category B pulses can be produced by a downward-propagating recoil leader.
As a whole, these findings demonstrate that detailed observations of upward lightning can offer valuable insight into the complex processes underlying lightning initiation and propagation.}

\keywords{preliminary breakdown pulses, lightning initiation, electric field measurements, S\"antis tower, upward lightning}



\maketitle

\section{Introduction}\label{sec:int}



The initiation and early development of lightning discharges remain among the most challenging and incompletely understood aspects of atmospheric electricity research (e.g., \cite{dwyer_physics_2014}). 
Despite significant advancements in observation techniques, the fundamental mechanisms underlying lightning initiation continue to generate considerable scientific debate.

Numerous studies have investigated the initiation of downward negative flashes using various measurement techniques, including electric field sensors, high-speed cameras (HSC), and very high frequency (VHF) interferometry (e.g., \cite{proctor_vhf_1988, nag_analysis_2009, marshall_percentage_2014}), in both tropical and temperate thunderstorms \citep{gomes_comparison_1998}. 
The seminal work by \cite{clarence_preliminary_1957} proposed a three-phase mechanism—Breakdown, Intermediate, and Leader (BIL)—that precedes the first return stroke in downward negative flashes. 
They hypothesised that the initial breakdown (Stage B) differs fundamentally from the stepped leader process (Stage L). 
This hypothesis has been both supported and challenged by subsequent research, with \cite{proctor_vhf_1988} suggesting, based on VHF observations, that similar discharge processes might occur in both the breakdown and leader stages.


One significant limitation in the study of downward lightning is the inability to directly measure the current waveforms associated with the initiation process. 
Unlike tower-initiated upward lightning, where direct current measurements are possible, researchers studying downward lightning must rely primarily on remotely-sensed electromagnetic fields.
HSC observations are also hampered by cloud opacity, whereas the initiation of upward flashes is frequently visible to optical measurements.

An earlier investigation by \cite{azadifar_similarity_2018} identified parallels between the initial stage of upward negative leaders (UNLs) and the preliminary breakdown phase in downward negative flashes. 
They observed two distinct types of electric field pulses associated with UNLs, which they labelled ``Category A'' and ``Category B'' pulses, respectively, remarkably similar to the ``Classical'' and ``Narrow'' preliminary breakdown pulses (PBPs) identified in downward negative flashes. 

In this study, we expand on their dataset and analysis, leveraging the measurement capabilities of the S\"antis Lightning Research Facility in Switzerland to analyse simultaneous data of channel-base current, vertical electric field, and (when available) HSC recordings of the initial stage of upward positive flashes (UPFs) initiated from the S\"antis Tower. 
By performing a more comprehensive comparative statistical analysis of these measurements with observations of PBPs recorded in downward negative leaders reported in the literature, we aim to establish whether these similarities are consistent with related physical processes, potentially providing new insights into the early stages of lightning leader development and possibly initiation.
Moreover, the inclusion of current pulse characteristics and HSC recordings in this expanded analysis allows us to validate existing physical models for PBPs in the context of UNL pulses, further supporting the notion of a shared origin.

The paper is organised as follows: 
Section~\ref{sec:method} details the methodology employed in this study;
Section~\ref{sec:meas} presents the measurement results, including the different pulse type classifications and their statistical characteristics; 
Section~\ref{sec:resu} provides a comparative analysis between leader pulses in upward lightning and PBPs in downward flashes, and examines the validity of existing models; and 
Section~\ref{sec:sumcon} summarises our key findings and discusses their implications for lightning research.


\section{Methods}\label{sec:method}


\begin{figure}[h]
    \centering
    \includegraphics[width=\textwidth]{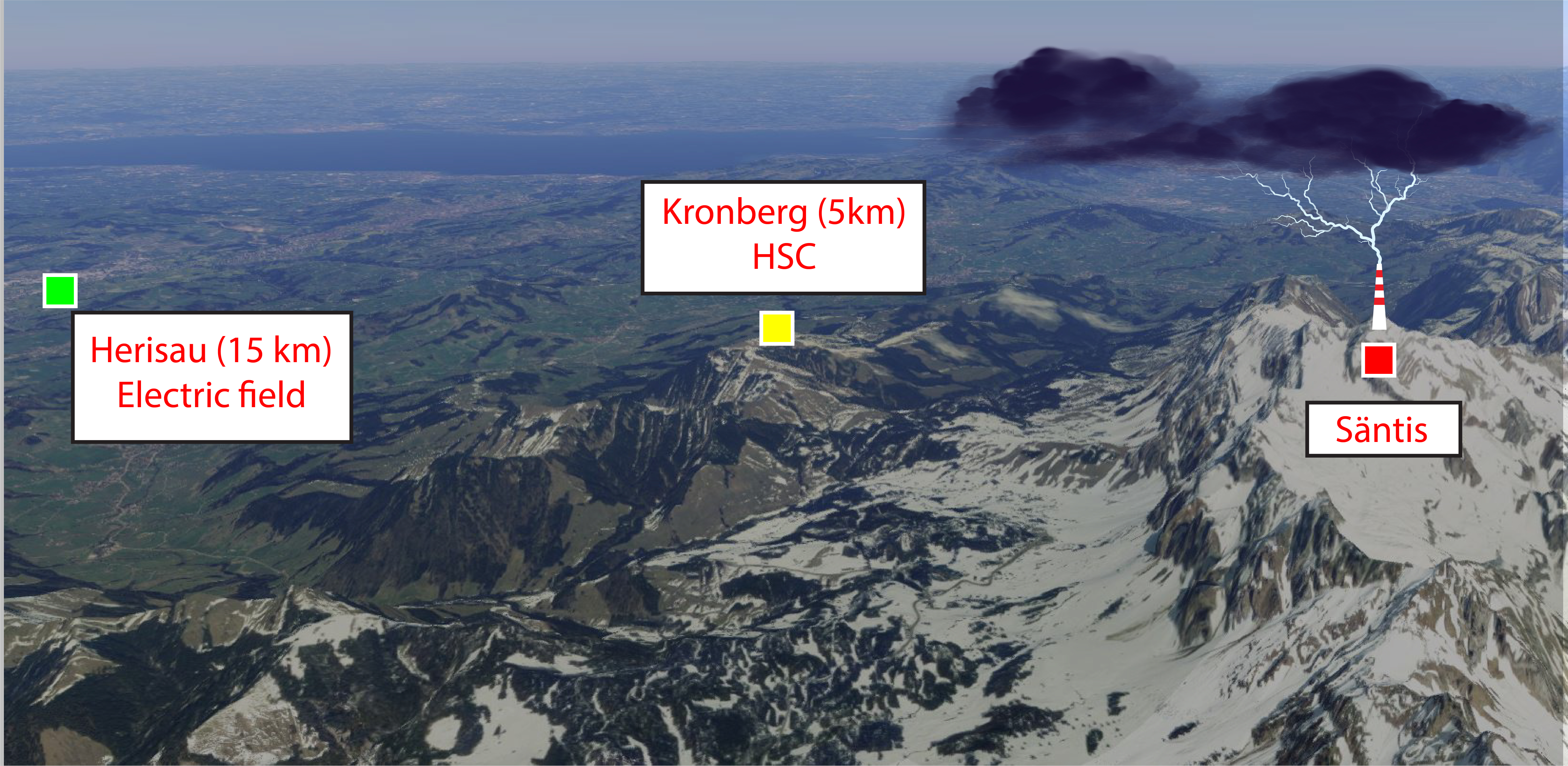}
    \caption{Depiction of the Mt. S\"antis research stations relevant to this study.}\label{fig:santis}
\end{figure}


The Mt. S\"antis Lightning Research Facility, shown in Figure~\ref{fig:santis}, is situated at 2502~m ASL in the Appenzell Alps of north-eastern Switzerland, and experiences $>$100 direct lightning strikes per year to its 124-meter-tall tower, which is equipped with a comprehensive current measurement system consisting of Rogowski coil and B-dot sensor pairs at two different heights: 24 and 82 meters above ground level (AGL). The system has a sampling rate of 50~MHz.
Five kilometers away, atop Mt. Kr\"onberg (1663 m ASL), is a high-speed camera (HSC) operating at 24,000 fps, with an exposure time of $\sim 41~\mu$s and a resolution of 512$\times$512 pixels.
Fifteen kilometres away, atop a building in Herisau, Switzerland (771~m ASL) with line-of-site, lies an flat-plate electric field antenna with a frequency range of 40~Hz to 40~MHz, a sampling rate of 10~MHz, and time constants of 4.2~ms (in 2014) and 8.4~ms (2021).
These settings are such that the sub-millisecond pulse temporal parameters measured herein are unaffected; furthermore, we do not expect significant signal propagation effects on these time scales at this distance.
This E-field signal is synchronised with the tower current signal by GPS timestamp if the antennae are functional at the time of the flash. 
If not, manual synchronisation can be carried out via waveform matching, with an uncertainty of $\sim 1/3 ~\mu$s. More detailed information on the S\"antis measurement system can be found in \cite{rachidi_santis_2022}.

All computational data analysis and presentation were carried out using MATLAB and the Python programming languages, in particular the NumPy, Pandas, SciPy, and Matplotlib libraries.


\section{Measurements}\label{sec:meas}

\subsection{Experimental Data}

Our dataset consists of observations from ten UPFs (eight Type 2 and two Type 1 according to the classification of \cite{romero_positive_2013}) initiated from the S\"antis tower between 2014 and 2021. 
For each flash, we collected current measurements in the Tower, vertical electric field measurements at a distance of 14.7~km, and, in one case, high-speed video footage. 

Table~\ref{tab:flashes} provides detailed information about each flash, including their identifier, timestamp, and, in addition to the specific data available, whether or not there was any prior lightning activity in a 30 km radius in the 3 seconds prior to initiation (as confirmed by the EUCLID Lightning Location System; see \cite{smorgonskiy_analysis_2015}). 
We included the information on prior lightning activity as previous studies have demonstrated the impact of preceding nearby lightning on upward leader formation (e.g., \cite{wang_observed_2008, sunjerga_initiation_2021}).
See Section~\ref{sec:method} for more methodology details.


\begin{table}
\centering
\caption{Upward Positive Lightning Flashes Analysed -- Data Summary}\label{tab:flashes}
\begin{tabular}{ c | m{5em} | c | c | c | c } 
\toprule
\textbf{Flash} & \textbf{Date \newline UTC} & \multirow[t]{2}{*}{\makecell{Prior \\ Activity\footnotemark[1]}} & \multirow[t]{2}{*}{\makecell{Current \\ (\& derivative)}} & \multirow[t]{2}{*}{\makecell{High-speed \\ camera}} & \multirow[t]{2}{*}{\makecell{E-field \\ (15-km)}} \\ 
\midrule
1 & 2014-10-21 \newline 21:24:27 & NO & YES & NO & YES \\ \hline 
2 & 2014-10-21 \newline 22:53:21 & NO & YES & NO & YES \\ \hline 
3 & 2014-10-21 \newline 22:56:47 & NO & YES & NO & YES \\ \hline 
4 & 2014-10-21 \newline 22:59:39 & NO & YES & NO & YES \\ \hline 
5 & 2014-10-21 \newline 23:12:22 & NO & YES & NO & YES \\ \hline 
6 & 2021-06-28 \newline 23:08:40 & YES & YES & NO & YES \\ \hline 
7 & 2021-07-24 \newline 16:06:07 & NO & YES & NO & YES \\ \hline 
8 & 2021-07-24 \newline 16:24:03 & NO & YES & YES & YES \\ \hline 
9 & 2021-07-30 \newline 18:00:10 & YES & YES & NO & YES \\\hline 
10 & 2021-07-30 \newline 18:04:53 & YES & YES & NO & YES \\ 
\botrule
\end{tabular}
\footnotetext[1]{in a 30~km radius in the 3 seconds prior to initiation.}
\end{table}


Type 1 and Type 2 UPFs, as defined by \cite{romero_positive_2013}, are distinguished by the presence (Type 1) or absence (Type 2) of a large unipolar return stroke-like current pulse following the upward negative stepped leader phase.
However, both types share the same physical mechanism and characteristics in their early stage, beginning with an UNL initiated at the tip of the strike object, which manifests as a waveform lasting on the order of 100 milliseconds, which is punctuated by large, oscillatory pulse trains (e.g., \cite{heidler_characteristics_2015, oregelchaumont_underlying_2025}).
These observations are further confirmed by the statistical similarity of the current and E-field pulse characteristics: with the exception of average current amplitudes being larger by a factor of $\sim2-3$ for Type 2 UPFs, all other calculated means fell within a standard deviation of each other for Type 1 and 2 UPFs.
As such, both types are grouped together in the following analysis.

It should be noted that flashes 7, 8 and 9 occurred during the Laser Lightning Rod project presented in \cite{houard_laser-guided_2023} (therein labelled L1, L2 and L3, respectively), while the laser was on.
The guiding effect discussed therein was observed over the first $\sim$50~m of propagation (see their Fig. 2), but with no clear evidence of laser-induced lightning \textit{initiation}.
The presence of the laser beam does not have any obvious effect on the pulse characteristics analysed in this study (i.e., no consistent differences with significant confidence observed), though we leave the door open to this possibility as the number of observed ``laser-guided'' flashes was too small to draw definitive conclusions.
See \cite{oregel-chaumont_direct_2024} for further discussion.

\subsection{Pulse Categories and Characteristics}

Through analysis of the simultaneous current and electric field measurements for over 70 pulses, we observe the two distinct categories of pulses associated with upward negative stepped leaders, as defined by \cite{azadifar_similarity_2018}:

\subsubsection{Category A Pulses}
Category A pulses are characterised by unipolar or bipolar vertical electric field signatures, the former typically negative and the latter typically with an initial positive half-cycle\footnote{Approximately 12\% exhibited inverted polarity with a negative initial half-cycle, a phenomenon also noted in the PBPs of downward negative leaders \citep{ogawa_initiation_1993}.}, correlated with negative unipolar current pulses at the tower.\footnote{We define a negative current as negative charge being transferred upward. For the E-field, we use the physics sign convention.}

Bipolar E-field pulses have an average duration on the order of $20-25~\mu$s and 10–90\% risetime on the order of $\sim 5~\mu$s, consistent with the characteristics of ``Classical'' PBPs observed in downward lightning.
Unipolar E-field pulses are noticeably shorter, with an average risetime and half-width of around 1.5 and 2.5 $\mu$s, respectively, comparable to values measured by \cite{wu_upward_2020}. It is worth noting, however, that we observed these pulses in only one flash (\#7 in Table~\ref{tab:flashes}), and have therefore been excluded from the primary analysis presented here until an expanded dataset can yield more reliable results.

\begin{figure}
\centering
    \begin{subfigure}[t]{0.49\textwidth}
        \centering
        \includegraphics[width=\textwidth]{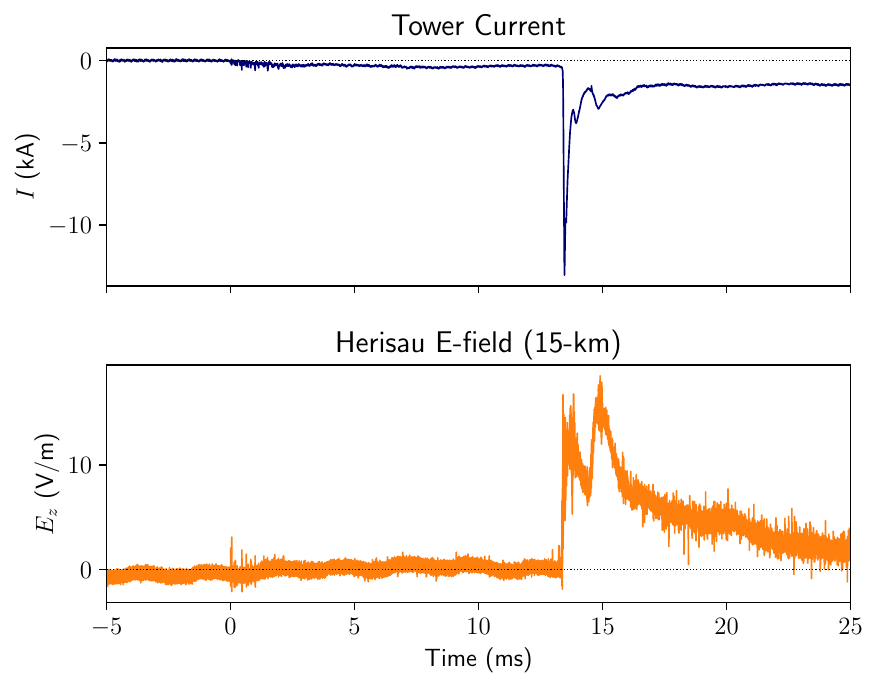}
        \subcaption{``Whole'' flash current and E-field waveforms associated with UPF \#8.}\label{fig:UP2wf}
    \end{subfigure}
    \hfill
    \begin{subfigure}[t]{0.49\textwidth}
        \centering
        \includegraphics[width=\textwidth]{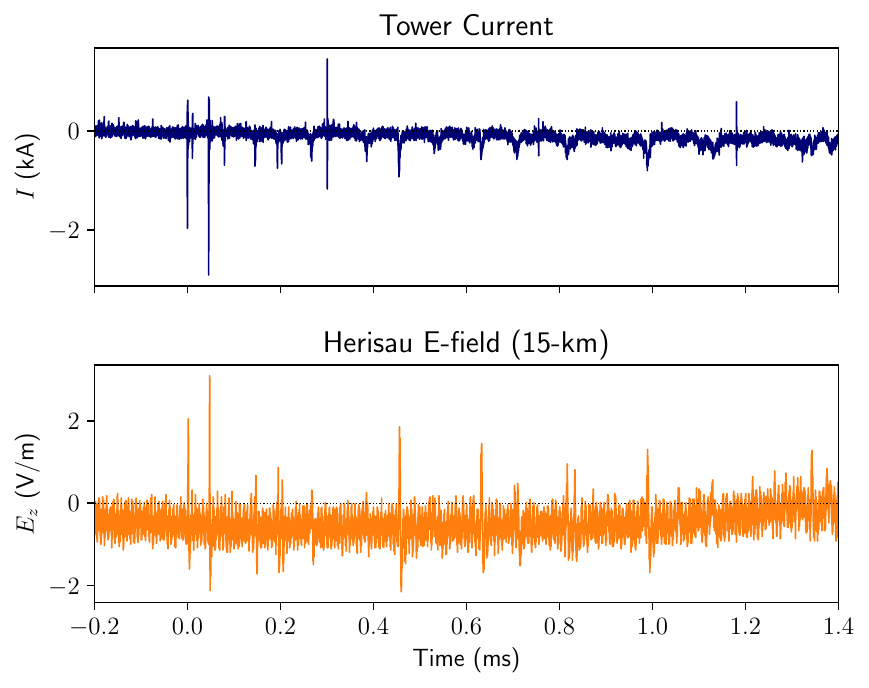}
        \subcaption{Typical Category A current and E-field pulse train.}\label{fig:bppt} 
    \end{subfigure}
    \begin{subfigure}[t]{0.49\textwidth}
        \centering
        \includegraphics[width=\textwidth]{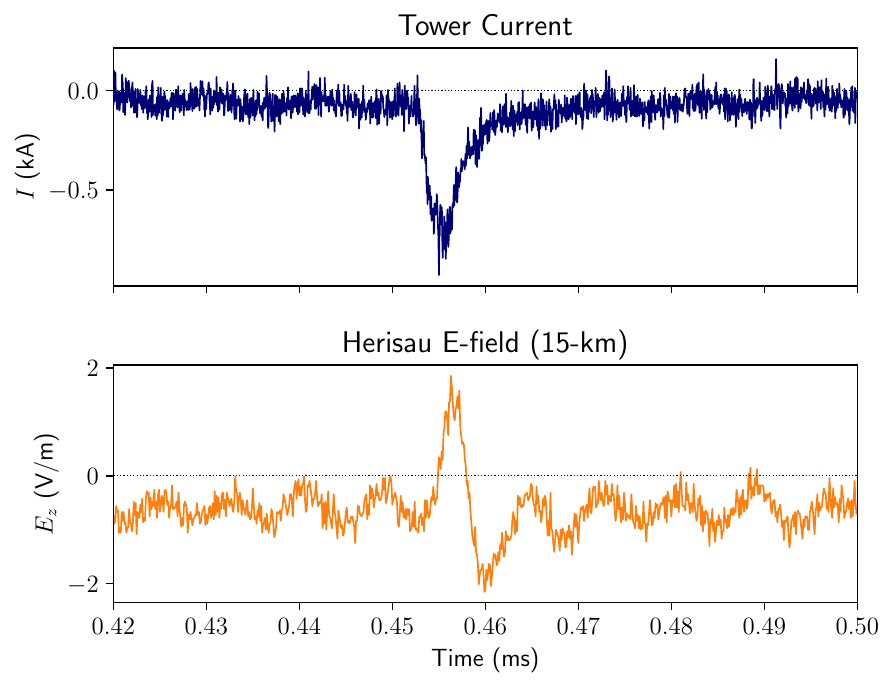}
        \subcaption{An individual bipolar Category A pulse.}\label{fig:bpps} 
    \end{subfigure}
    \hfill
    \begin{subfigure}[t]{0.49\textwidth}
        \centering
        \includegraphics[width=\textwidth]{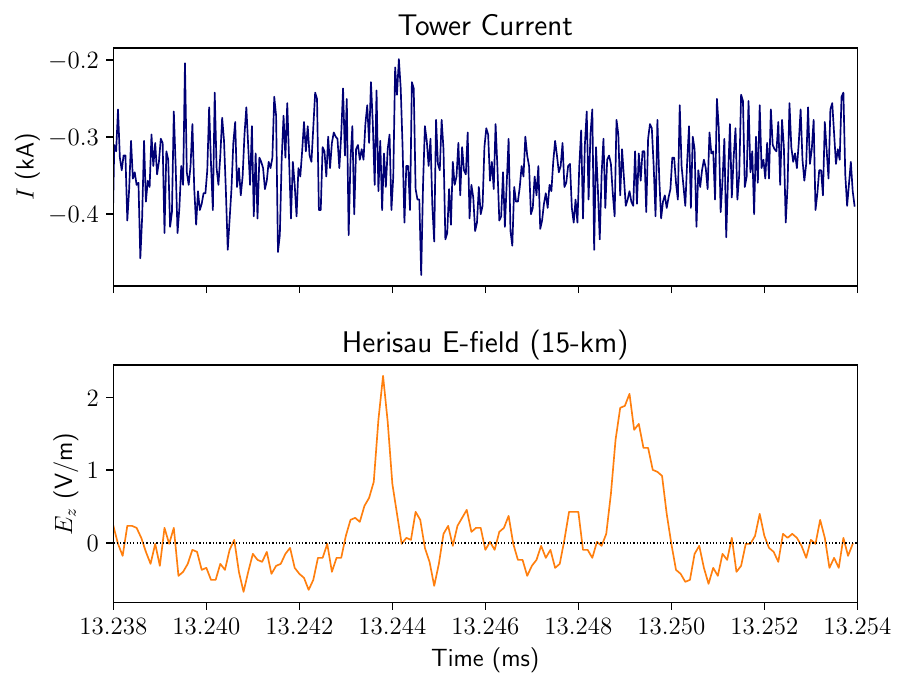}
        \subcaption{Two unipolar Category B pulses.}\label{fig:cbps1}
    \end{subfigure}
\caption{Current and  electric field waveforms associated with the upward positive flash \#8 listed in Table~\ref{tab:flashes}. (a) Overall waveforms. (b) A bipolar Category A pulse train occurring in the early stage of the ICC. (c) Current and E-field waveforms corresponding to an individual Category A pulse. (d) Two typical Category B pulses observed at the end of the ICC phase. The current in (a) has been wavelet-filtered to remove high-frequency electronic noise. The E-field waveforms follow the physics sign convention. Time is measured relative to the onset of the ICC. 
}\label{fig:CatApulses}
\end{figure}

Both unipolar and bipolar Category A electric field pulses are well correlated with negative unipolar initial continuous current (ICC) pulses associated with the stepping of the UNL.
Figure~\ref{fig:UP2wf} presents the overall current and  electric field waveforms associated with the upward positive flash \#8 listed in Table~\ref{tab:flashes}. 
The current waveform begins with an ICC, followed by the characteristic Type 1 main pulse approximately 13 milliseconds after the onset of the ICC.
Figure~\ref{fig:bppt} shows a bipolar Category A pulse train occurring in the early stage of the ICC.
Figure~\ref{fig:bpps} displays the current and E-field waveforms corresponding to a single bipolar Category A pulse. 


\subsubsection{Category B Pulses}
Category B pulses are characterised by unipolar (positive or negative) or bipolar E-field signatures with much shorter temporal characteristics than Category A pulses, exhibiting average half-widths and risetimes of less than $1~\mu$s. 
More importantly, these pulses lack correlation with any significant current pulses measured at the tower, and tend to appear later in leader development. 
Figure~\ref{fig:cbps1} illustrates two typical Category B pulses also belonging to flash \#8, showing their faster temporal characteristics compared to Category A pulses.
Note that the lack of current correlation for Category B pulses suggests a different physical mechanism than Category A; see Section~\ref{sec:narrowPhysMech} for further discussion.


\section{Results}\label{sec:resu}


\subsection{Statistical Analysis}

\begin{figure}
\centering
\begin{subfigure}[t]{0.47\textwidth}
    \centering
    \includegraphics[width=\textwidth]{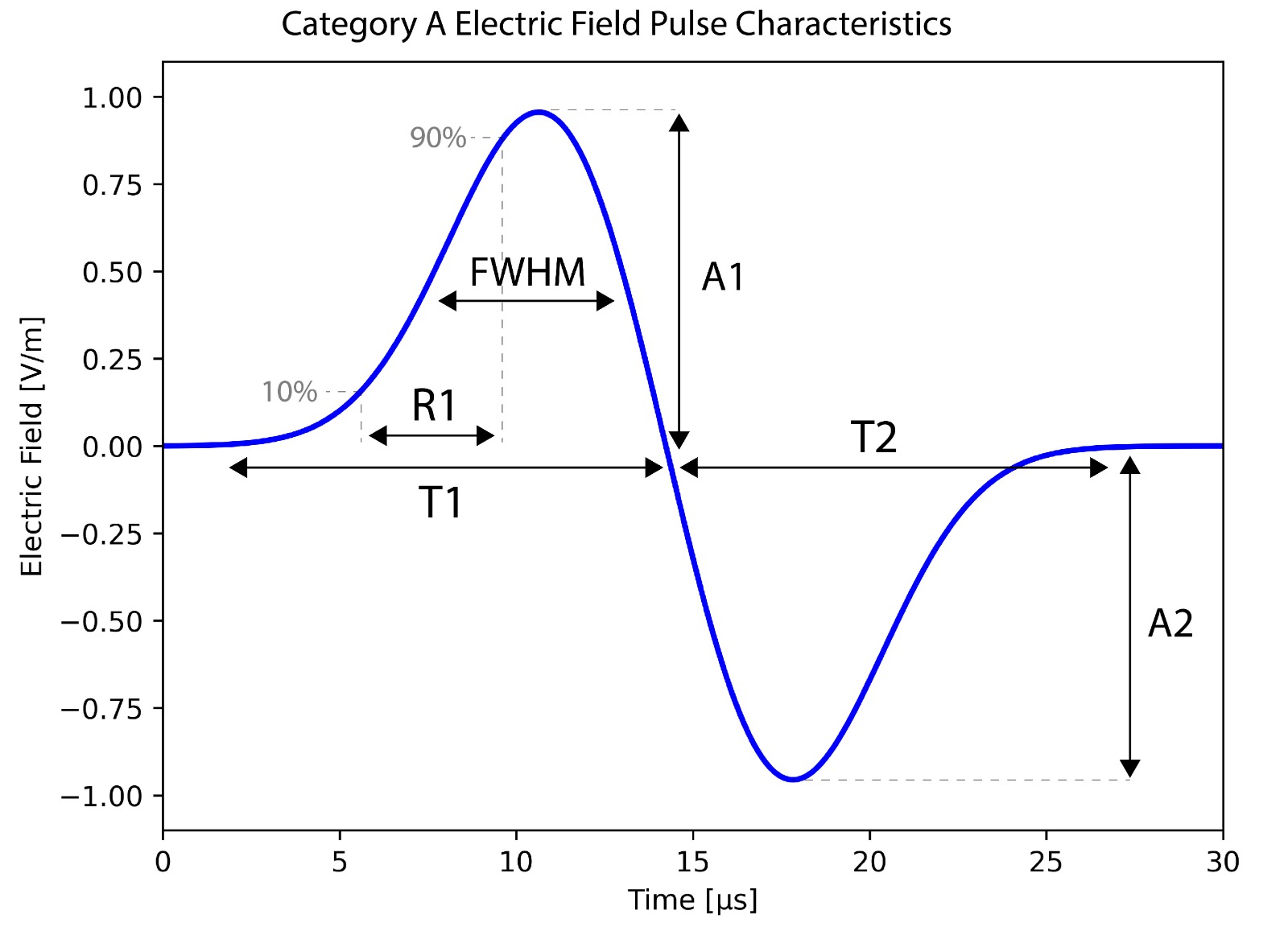}
    \end{subfigure}
\hfill
\begin{subfigure}[t]{0.52\textwidth}
    \centering
    \includegraphics[width=\textwidth]{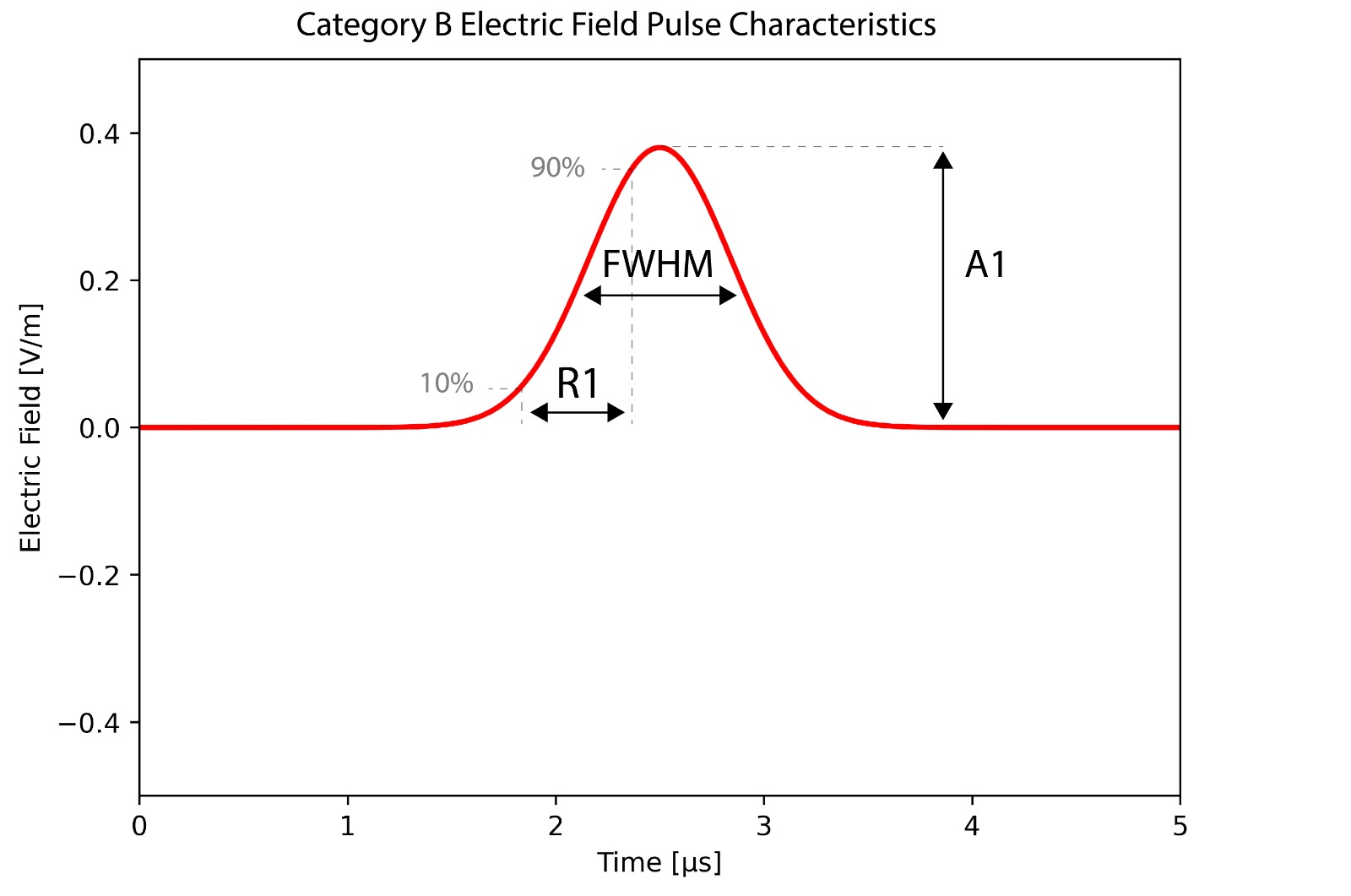}
\end{subfigure}
\caption{Characteristic measurements for the different pulses investigated in this study. Mean values have been selected for each characteristic.}\label{fig:pcs}
\end{figure}

We conducted detailed measurements of pulse characteristics as defined in Figure~\ref{fig:pcs}, analysing parameters such as the 10-90\% risetime, the full width at half maximum (FWHM)\footnote{Note that \cite{azadifar_similarity_2018} used the term half-peak beam width (HPBW) to refer to the same parameter.}, and for bipolar E-field pulses, the amplitudes and durations of both half-cycles.



\begin{table}[t]
\centering
\caption{Pulse characteristic statistics ($\mu_a \pm \sigma_a$ \textbar\, ${\mu_g}^{\mu_g(\sigma_g - 1)}_{\mu_g(1-1/\sigma_g)}$). Subscripts $I$ and $E$ denote current and E-field pulses, respectively. E-field pulse amplitudes are normalised to a distance of 100 km, assuming a $1/d$ distance dependence.}\label{tab:pulseStats} 
\begin{tabular}{ c | c | c } 
\toprule
\textbf{Pulse Type ($n$)} & \textbf{Category A} (65) & \textbf{Category B} (15) \\ 
\midrule
R$_I$ [$\mu$s] & 5.12$\pm$2.67 \textbar\, $4.49^{+3.09}_{-1.83}$ & --- \\ 
FWHM$_I$ [$\mu$s] & 6.61$\pm$2.07 \textbar\, $6.21^{+2.99}_{-2.02}$ & --- \\ 
A$_I$ [kA] & 2.19$\pm$1.61 \textbar\, $1.55^{+2.32}_{-0.93}$ & --- \\ 
$|dI/dt|_\mathrm{max}$ [kA/$\mu$s] & 3.5$\pm$2.7 \textbar\, $2.8^{+2.4}_{-1.3}$ & --- \\ \hline 
T1+T2 [$\mu$s] & 18.3$\pm$10.8 \textbar\, $15.9^{+10.9}_{-6.5}$ & --- \\ 
R1$_E$ [$\mu$s] & 4.3$\pm$3.3 \textbar\, $3.2^{+4.4}_{-1.9}$ & 0.7$\pm$0.5 \textbar\, $0.5^{+0.6}_{-0.3}$ \\ 
FWHM$_E$ [$\mu$s] & 3.9$\pm$2.8 \textbar\, $3.1^{+3.3}_{-1.6}$ & 0.7$\pm$0.4 \textbar\, $0.6^{+0.4}_{-0.3}$ \\ 
A1$_E$ [V/m] & 1.01$\pm$0.61 \textbar\, $0.80^{+0.87}_{-0.42}$ & 0.35$\pm$0.27 \textbar\, $0.28^{+0.26}_{-0.13}$ \\ 
$|dE/dt|_\mathrm{max}$ [V/m/$\mu$s] & 13.4$\pm$7.0 \textbar\, $11.3^{+9.8}_{-5.3}$ & 7.37$\pm$4.59 \textbar\, $6.26^{+4.67}_{-2.67}$ \\ \hline 
Approx. 2D $l_S$ [m]\footnotemark[1] & 26$\pm$8 \textbar\, $24^{+9}_{-7}$ & $>56 \pm 4$ \\ 
\botrule
\end{tabular}
\footnotetext[1]{Only one flash provided these data; see Table~\ref{tab:UP2pulses}. The average $l_S$ for the much faster Cat. B pulse train was obtained by dividing the measured branch length by the number of identified pulses.}
\end{table}


Table~\ref{tab:pulseStats} summarises the statistical properties of both pulse categories, based on 65 Category A pulses and 15 Category B pulses.
$\mu_a \pm \sigma_a$ and ${\mu_g}^{\mu_g(\sigma_g - 1)}_{\mu_g(1-1/\sigma_g)}$ refer to the arithmetic and geometric means and standard deviations, assuming Gaussian and log-normal distributions, respectively.
In addition to the average risetimes of $4.7~\mu$s (AM) and $3.7~\mu$s (GM), 
bipolar Category A electric field pulses were found to have half-widths of $4.2~\mu$s (AM) and $3.3~\mu$s (GM) 
and peak amplitudes (A1$_E$, normalised to 100~km) of $\sim$1.0~V/m (AM) and 0.8~V/m (GM). 
The corresponding current pulses had risetimes of $\sim 5.3~\mu$s (AM) and $4.5~\mu$s (GM), half-widths of $\sim 6.6~\mu$s (AM) and $6.1~\mu$s (GM), and peak currents of $\sim$2.2~kA (AM) and 1.5~kA (GM). 
The mean step lengths $l_S$ are based on HSC observations of the nine Category A pulses and twelve Category B pulses measured in Flash \#8.
These were estimated by measuring the 2D pixel length of the channel in question, and converting to meters using the tower as a reference (124~m / 37~pix). 
Note that Category B step lengths are more than twice those of Category A pulses, and in both cases represent a lower bound on the true 3D length.
This points to a difference in physical mechanism: 
Category A pulses are associated with the stepping of the UNL, i.e., virgin breakdown, while Category B pulses were observed to be due to a recoil leader retracing a pre-existing channel: see Section~\ref{sec:narrowPhysMech} for further discussion.

\begin{figure}
\centering
    \begin{subfigure}[t]{0.53\textwidth}
        \centering
        \includegraphics[width=\textwidth]{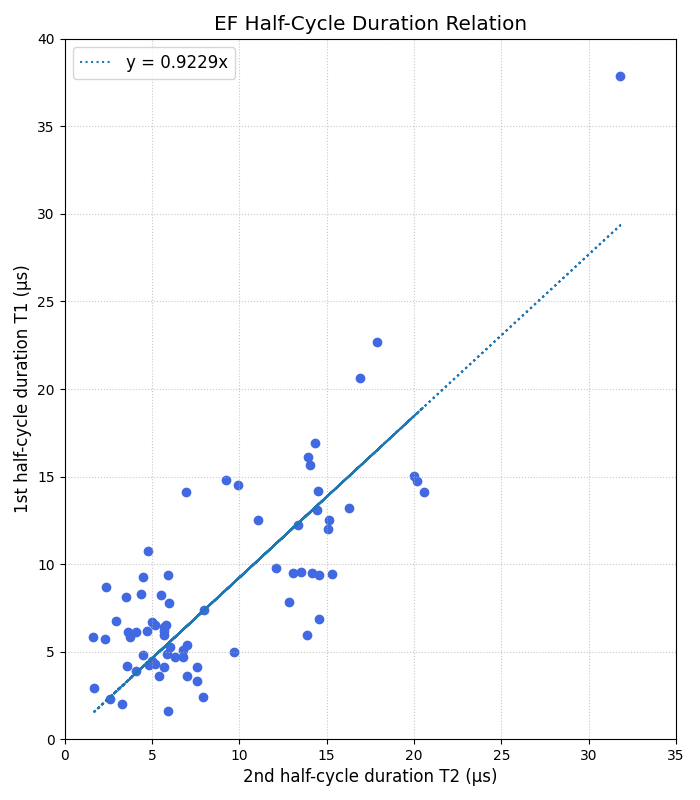}
        \subcaption*{$R^2=0.86$}
    \end{subfigure}
    \hfill
    \begin{subfigure}[t]{0.46\textwidth}
        \centering
        \includegraphics[width=\textwidth]{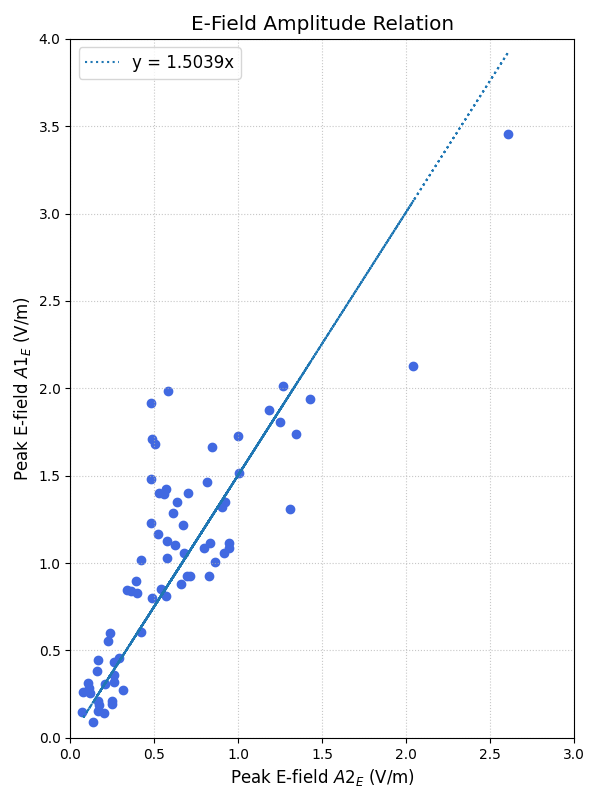}
        \subcaption*{$R^2=0.90$}
    \end{subfigure}
\caption{Plots depicting the relationships between the primary half-cycle characteristics of bipolar Category A pulses.}
\label{fig:CatA-HC-rels}
\end{figure}

Notably, we observed time-dependent characteristics in the bipolar pulse trains, with increased peak currents ($\rho = 0.71$, $r_s = 0.80$, $\tau = 0.58$) and decreased maximum E-field derivatives ($\rho = -0.75$, $r_s = -0.69$, $\tau = -0.49$) as the flash progressed,\footnote{$\rho$, $r_s$ and $\tau$ represent the Pearson, Spearman and Kendall correlation coefficients, respectively, calculated from the aforementioned sample size of $n=65$.} as shown in Figure~\ref{fig:bppt} and Table~\ref{tab:UP2pulses} ($t_{SL}$ represents the time from initiation of the stepped leader, i.e., the onset of the ICC).
In the last two columns we can see how the step length $l_S$ (and its vertical component $\Delta z$) also grow larger.
We also found significant correlations between various Category A pulse parameters, including linear relationships between the temporal widths of the initial and second half-cycles ($\rho = 0.79$, $r_s = 0.61$, $\tau = 0.40$) 
and between their peak amplitudes ($\rho = 0.83$, $r_s = 0.78$, $\tau = 0.60$);
see Figure~\ref{fig:CatA-HC-rels}.


Category B pulses demonstrated a very strong linear relationship between their amplitude and maximum derivative ($\rho = 0.94$, $r_s = 0.84$, $\tau = 0.73$), yielding a minimum characteristic risetime of A$_E$/$|dE/dt|_\mathrm{max}$ = 0.3~$\mu$s (with a coefficient of determination $R^2 = 0.95$), about 1/2 the measured risetime R1$_E$, though a larger dataset is needed to confirm this last relationship, due to the relatively small sample size of $n=15$.



\begin{sidewaystable}
\centering
\caption{Characteristics of bipolar pulses observed in Flash \#8}\label{tab:UP2pulses}
\begin{tabular}{ c | c | c | c | c | c | c | c | c | c | c }
\toprule
$t_{SL}$ [$\mu$s] & R$_I$ [$\mu$s] & FWHM$_I$ [$\mu$s] & A$_I$ [kA] & $|\frac{dI}{dt}|_\mathrm{m}$ [$\frac{\mathrm{kA}}{\mu\mathrm{s}}$] & R1$_E$ [$\mu$s] & FWHM$_E$ [$\mu$s] & A1$_E$ [$\frac{\mathrm{V}}{\mathrm{m}}$] & $|\frac{dE}{dt}|_\mathrm{m}$ [$\frac{\mathrm{V}}{\mathrm{m}\cdot\mu\mathrm{s}}$] & $l_S$ [m] & $\Delta z$ [m] \\
\midrule
  0 & 3.48$\pm$0.02 & 5.48$\pm$0.02 & 1.92$\pm$0.05 & 9.3$\pm$0.3 & 2.4$\pm$0.1 & 0.9$\pm$0.1 & 0.40$\pm$0.05 & 5.9$\pm$0.5 & 14$\pm$4 & 10$\pm$3 \\ 
 46 &  3.38 &  5.88 & 3.00 & 8.0 & 1.1 & 0.8 & 0.55 & 9.2 & 17 & 10 \\ 
145 & 11.78 &  8.38 & 0.64 & 2.9 & 0.5 & 0.9 & 0.14 & 5.1 & 22 & 20 \\ 
267 &  3.98 &  5.88 & 0.58 & 1.2 & 2.9 & 2.4 & 0.11 & 3.4 & 22 & 20 \\ 
455 &  3.88 &  6.38 & 0.90 & 1.1 & 1.4 & 1.8 & 0.23 & 3.8 & 24 & 20 \\
631 &  4.38 &  7.18 & 0.48 & 1.1 & 1.0 & 1.8 & 0.17 & 4.9 & 36 & 27 \\
817 &  4.48 & 14.18 & 0.44 & 0.7 & 2.0 & 1.3 & 0.09 & 3.7 & 34 & 20 \\
989 & 10.98 &  8.08 & 0.60 & 0.9 & 2.1 & 7.8 & 0.15 & 3.2 & 36 \textbar\, 31 & 34 \textbar\, 20 \footnotemark[1] \\ \hline
$\mu_a \pm \sigma_a$ & 5.67$\pm$3.06 & 7.68$\pm$2.65 & 1.07$\pm$0.86 & 3.1$\pm$3.2 & 1.7$\pm$0.8 & 2.2$\pm$2.2 & 0.23$\pm$0.15 & 6.0$\pm$1.5 & 26$\pm$8 & 20$\pm$7 \\
\bottomrule
\end{tabular}
\footnotetext[1]{This ``step'' exhibited simultaneous extension of both branches of the plasma channel.}
\end{sidewaystable}





\subsection{Similarity with PBPs in downward flashes}

\subsubsection{Overview of PBPs in Downward Lightning}

Preliminary breakdown pulses (PBPs) in downward negative flashes mark the initial stages of leader development and are traditionally classified into two main categories, presented in Figure~\ref{fig:cnpbps}, based on their waveform characteristics:
\begin{itemize}
    \item ``Classical'' PBPs: These typically feature bipolar E-field signatures with durations in the range of 20-40~$\mu$s. 
    They are predominantly observed during the initial Breakdown (B) stage of the BIL process.
    \item ``Narrow'' PBPs: These have much shorter durations ($< 2 ~\mu$s) and can be unipolar or bipolar (Nag and Rakov, 2008, 2009).
    They are commonly observed during the Leader (L) stage but can also appear superimposed on Classical PBPs during the B stage.
\end{itemize}
The physical mechanisms underlying these pulses remain contested. 
Using high-speed video footage, \cite{stolzenburg_luminosity_2013} proposed an ``initial leader" concept for Classical PBPs (in stage B) that differs from normal stepped leaders (in stage L), involving a dim linear feature moving downward, followed by impulsive breakdown at the lower end and an upward-moving brightness.
They hypothesised that the initial leader pulses stop occurring because the previous initial leaders transferred enough charge to reduce the ambient electric field near the initiation point.
In contrast, \cite{campos_visible_2013} observed that channel extension in stage B appears similar to ordinary leader extension in stage L, suggesting comparable mechanisms.

\cite{petersen_high-speed_2015} observed a bimodal distribution of the stepping process, involving both long (200+~m length) and short (10+~m length) steps. 
The same kind of pulses observed in stage L (``Narrow'' PBPs) were observed in stage B, superimposed on slower, ``Classical'' PB pulses. 
They suggested the presence of distant space leaders with their negative ends stepping downward and generating narrow pulses (like those in stage L), while their positive ends generate classical pulses when they attach to the previous main leader channel (major pulses in stage B).
With the ambient electric field reduced by the descent of the leader, less distant space leaders occur, which leads to the generation of only narrow pulses shortly before the return stroke, as the leader approaches the ground.

\subsubsection{Comparative Analysis of Pulse Characteristics}\label{sec:companal}


\begin{figure}[h]
\centering
    \begin{subfigure}[t]{\textwidth}
        \centering
        \includegraphics[width=\textwidth]{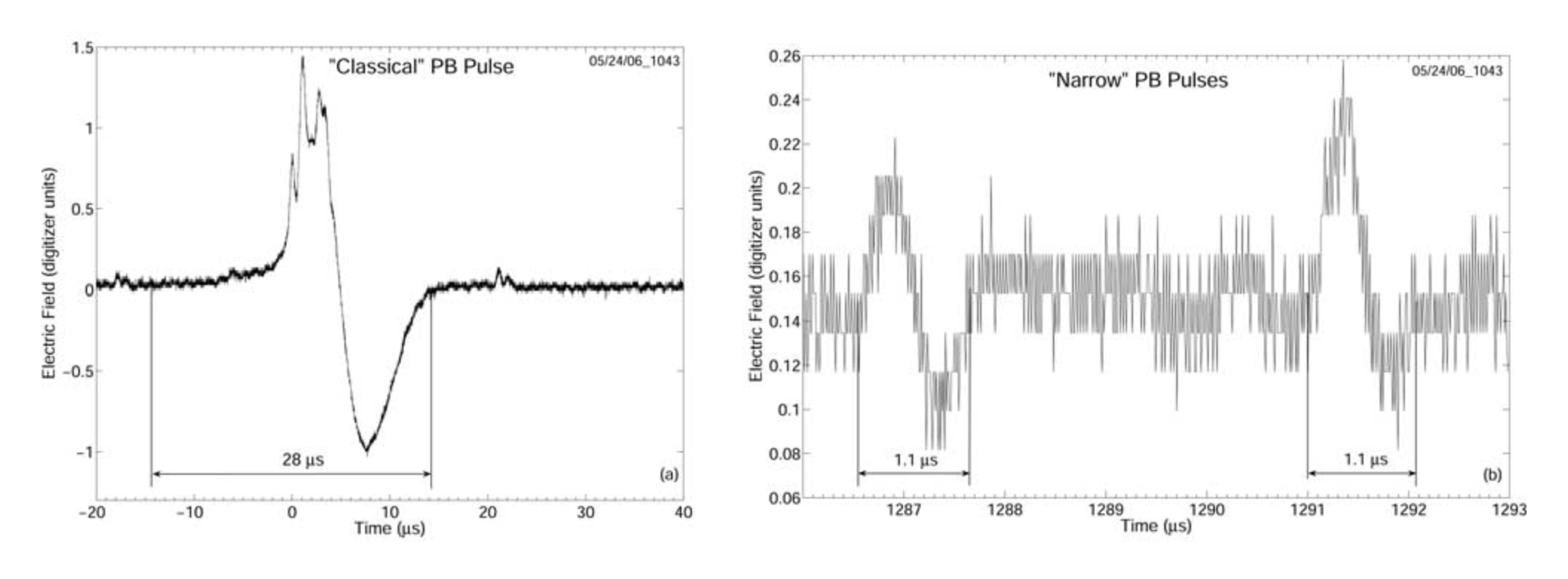}
        \subcaption{Sample ``Classical'' and ``Narrow'' PB pulse waveforms. Reprinted from \cite{nag_pulse_2008} (their Figure 7).}\label{fig:cnpbps}
    \end{subfigure}
    \begin{subfigure}[t]{0.49\textwidth}
        \centering
        \includegraphics[width=\textwidth]{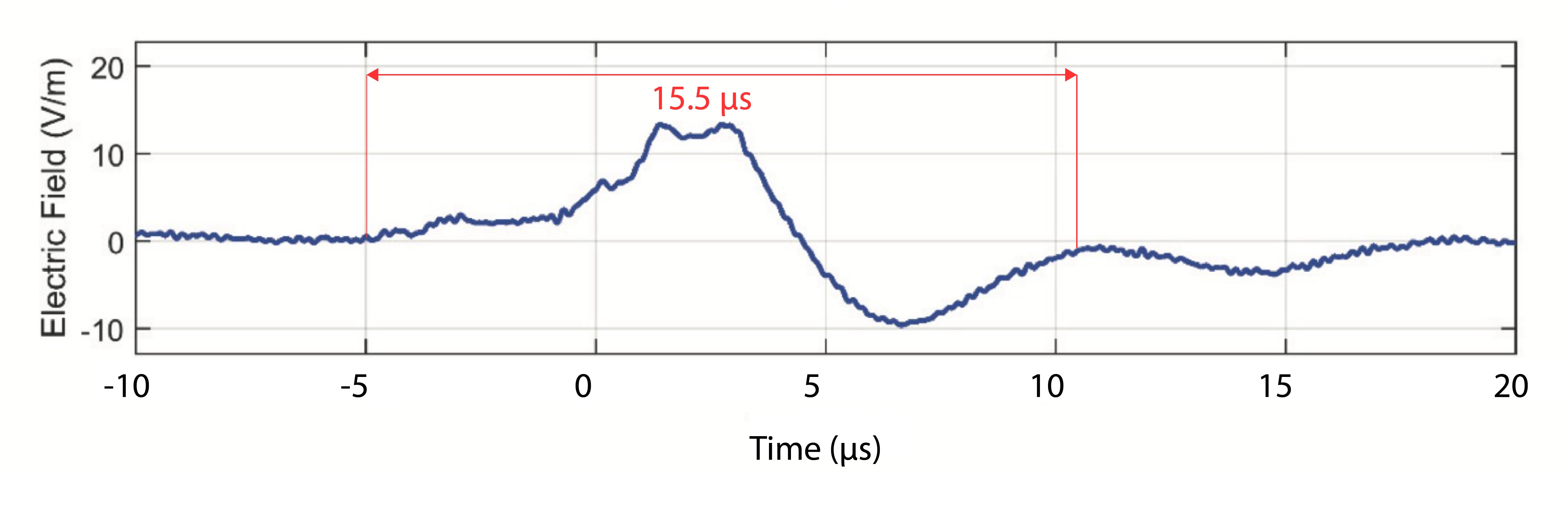}
        \subcaption{Typical Category A pulse waveform belonging to measured flash \#3.}
    \end{subfigure}
    \hfill
    \begin{subfigure}[t]{0.49\textwidth}
        \centering
        \includegraphics[width=\textwidth]{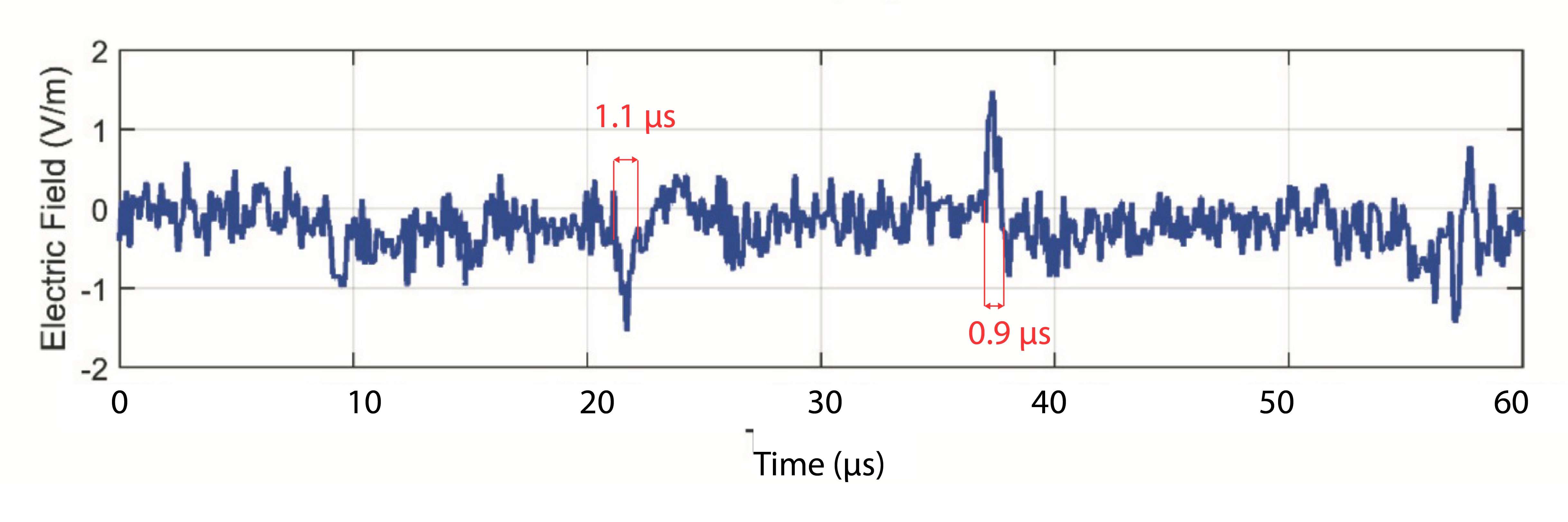}
        \subcaption{Typical Category B pulse waveform belonging to measured flash \#3.}
    \end{subfigure}
\caption{A visual illustration comparing ``Classical'' and ``Narrow'' PBP with  Category A and B pulse E-field waveforms, to emphasise their similarities. Plots (b) and (c) are adapted from \cite{azadifar_similarity_2018} (their Figures 2 \& 3, respectively).}\label{fig:PBPvCatAB}
\end{figure}


Our statistical analysis reveals notable similarities between bipolar Category A pulses in upward lightning and Classical PBPs in downward lightning (see a visual illustration presented in Figure~\ref{fig:PBPvCatAB}), particularly regarding key characteristic timescales:
\begin{itemize}
    \item Risetime (R1): The arithmetic mean and standard deviation of the  10-90\% risetime are $4.6 \pm 3.4~\mu$s for Category A pulses, which aligns\footnote{We infer relative statistical ``alignment'' as each mean being within the other’s standard deviation.} with values of $6.8 \pm 5.5~\mu$s reported for Classical PBPs in a recent study by \cite{shi_characterization_2024} on intracloud (IC) flashes;
    \item Full width half max (FWHM): Our measured average of $4.1 \pm 3.0~\mu$s also matches \cite{shi_characterization_2024}'s $3.8 \pm 2.2~\mu$s;
    \item Zero-crossing Time (T1): The average first half-cycle duration of $9.0 \pm 5.9~\mu$s, on par with \cite{wu_preliminary_2013}'s $6.1~\mu$s, though seemingly smaller than the $26.2~\mu$s measured by \cite{shi_characterization_2024}, actually fits nicely within their standard deviation of $21.1~\mu$s;
    \item Pulse Duration (T1+T2): Bipolar Category A pulses have an average duration of $18.0 \pm 11.3~\mu$s, falling within the range of $\sim$5 to 100+~$\mu$s (with an average of $\sim 20~\mu$s) reported for Classical PBPs in negative cloud-to-ground (CG) flashes \citep{wang_beijing_2016, zhu_study_2016, granados_characterization_2022}.
\end{itemize} 
This statistical comparison to recent literature is presented in Table~\ref{tab:litcom} 
(See Table 3 of \cite{sekehravani_preliminary_2025} for more details).
The inverted IC flashes of \cite{shi_characterization_2024} have been included for their qualitative (and statistical) similarities, including “downward PB propagation”, a negative initial E-field change, and comparable temporal characteristics (as we show).
Similarly, Category B pulses share temporal characteristics with Narrow PBPs, with typical durations an order of magnitude shorter than Category A/Classical PBPs. 
\cite{nag_pulse_2008}, 2009 classified them as $<4~\mu$s, and our sub-microsecond measured risetimes and half-widths depicted in Table~\ref{tab:pulseStats} are consistent with this.
A visual comparison of Category A/B and Classical/Narrow PB pulses is presented in Figure~\ref{fig:PBPvCatAB}.



\begingroup
\renewcommand{\thefootnote}{\alph{footnote}} \begin{table}
\centering
\caption{Characteristics of classical PBPs in negative flashes reported in the literature, compared with 65 Category A pulses in upward positive flashes observed at S\"antis. (AM \textbar\, GM)}\label{tab:litcom}
\begin{tabular}{ c | c | c }
\toprule
Pulse Characteristic & PBPs [literature] & Bip. Cat.A pulses [this study] \\
\midrule
Risetime (R1$_E$) [$\mu$s] & 6.8 \textbar\, 5.2 \footnotemark[1] & 4.6 \textbar\, 3.7 \\ \hline
FWHM [$\mu$s] & 3.8 \textbar\, 3.4 \footnotemark[1] & 4.1 \textbar\, 3.2 \\ \hline
\multirow{2}{*}{T1 [$\mu$s]} & 26.2 \textbar\, 19.9 \footnotemark[1] & 9.0 \textbar\, 7.5 \\ 
                             & 6.1 \textbar\, --- \footnotemark[2] & \\ \hline
\multirow{4}{*}{Duration (T1+T2) [$\mu$s]} & 11 \textbar\, --- \footnotemark[3] & \\
                                           & 26.2 \textbar\, 21.6 \footnotemark[4] & 18.0 \textbar\, 15.3 \\                                   
                                           & 25 \textbar\, 21 \footnotemark[5] & \\
                                           & 17.2 \textbar\, 11.9 \footnotemark[6] & \\
\botrule
\end{tabular}
\footnotetext[a]{\cite{shi_characterization_2024}}
\footnotetext[b]{\cite{wu_preliminary_2013}}
\footnotetext[c]{\cite{baharudin_comparative_2012}}
\footnotetext[d]{\cite{wang_beijing_2016}}
\footnotetext[e]{\cite{zhu_study_2016}}
\footnotetext[f]{\cite{granados_characterization_2022}}
\end{table}
\endgroup
 
The observation of inverted polarity in approximately 10\% of the bipolar Category A pulses (negative initial half-cycle)\footnote{These seven pulses were excluded from the primary statistical analysis due to their relative rarity.} also parallels similar findings in downward lightning studies by \cite{ogawa_initiation_1993}, further supporting the connection between these phenomena.

\begin{figure}
\centering
    \begin{subfigure}[t]{0.48\textwidth}
        \centering
        \includegraphics[width=\textwidth]{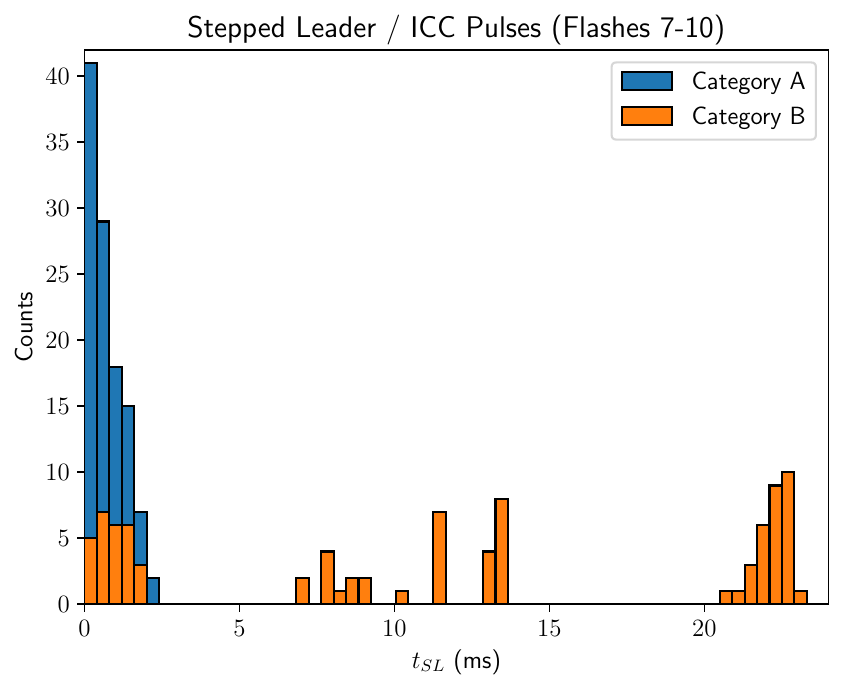}
        \subcaption{Pulse timing until $t_{SL}=24$ ms with $400~\mu$s bins.}
    \end{subfigure}
    \hfill
    \begin{subfigure}[t]{0.48\textwidth}
        \centering
        \includegraphics[width=\textwidth]{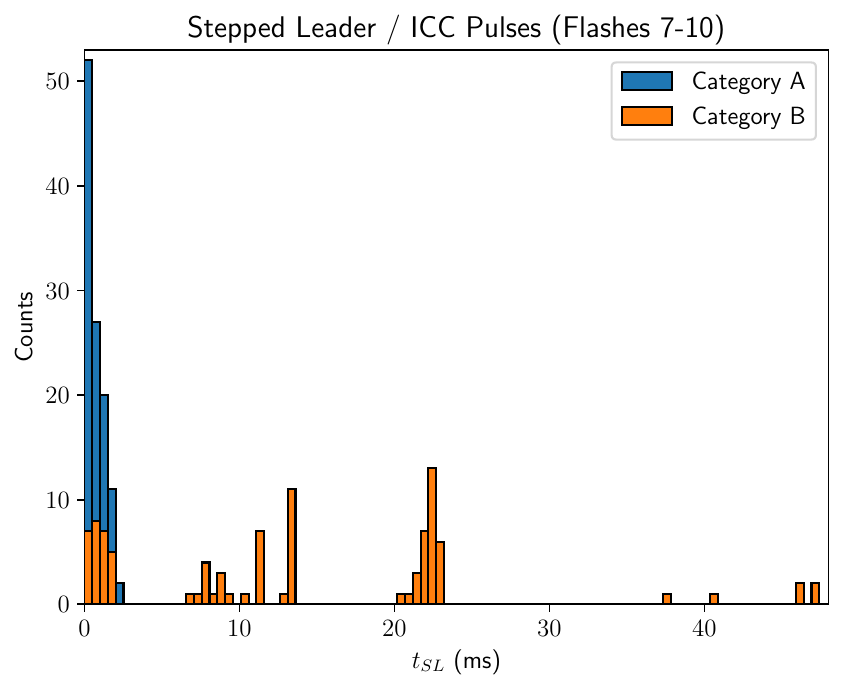}
        \subcaption{Pulse timing until $t_{SL}=48$ ms with $500~\mu$s bins.}
    \end{subfigure}
\caption{Histograms depicting the bimodal temporal distribution of Category A \& B pulses observed in Flashes \#7 through \#10.\protect\footnotemark \, The Type 1 UPF main pulses of flashes \#8 and \#10 occur at $t_{SL}=13.46$ and 8.94~ms, respectively. Note the quiet I phase in between the B and L phases dominated by Category A and B pulses respectively.}
\label{fig:hist} 
\end{figure}

Interestingly, we observed both Category A and B pulses occurring simultaneously during the early stage of leader development, followed by a later stage dominated exclusively by Category B pulses. 
This pattern parallels observations in downward negative flashes reported by \cite{petersen_high-speed_2015}, in which, as discussed in the previous section, they noted a bimodal distribution of the stepping process involving both long (200+~m) and short (10+~m) steps associated with the B and L phases, respectively (see their figures 1 \& 6).
The bimodal distribution of stepping processes observed in our data--with both Category A and B pulses present initially, followed by predominantly Category B pulses--mirrors the pattern described by \cite{petersen_high-speed_2015} for downward negative flashes, as shown in Figure~\ref{fig:hist}.
This suggests that similar space leader formation and attachment processes may be occurring in both upward and downward leader development.



\footnotetext{No Category B pulses were observed in Flash \#6.}

It is important to note, however, that the measured 2D step lengths for Classical PBPs are an order of magnitude larger than those measured for Category A pulses ($\sim$25~m), while our estimated Category B step lengths ($\sim$55~m) are of the same order as those reported during the L phase (Narrow PBPs), albeit a bit longer \citep{petersen_high-speed_2015}.
The former discrepancy is likely due to a difference in electric field geometry at the initiation point (in-cloud vs. tower tip), while the latter similarity is possibly due to the physical correspondence of downward stepping leaders approaching the ground (albeit of different polarities).


\begin{figure}
\centering
    \begin{subfigure}[t]{0.41\textwidth}
        \centering
        \includegraphics[width=\textwidth]{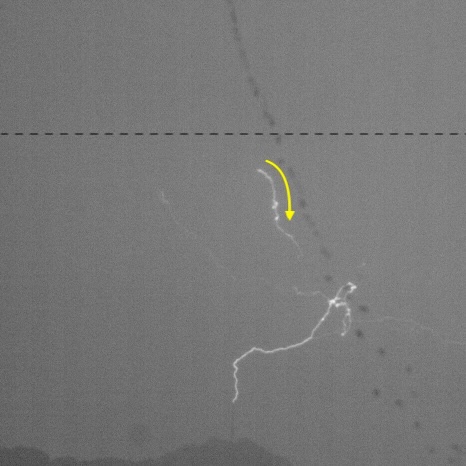}
        \subcaption{HSC frame containing the Category B pulses to the right.}\label{fig:f972}
    \end{subfigure}
    \hfill
    \begin{subfigure}[t]{0.57\textwidth}
        \centering
        \includegraphics[width=\textwidth]{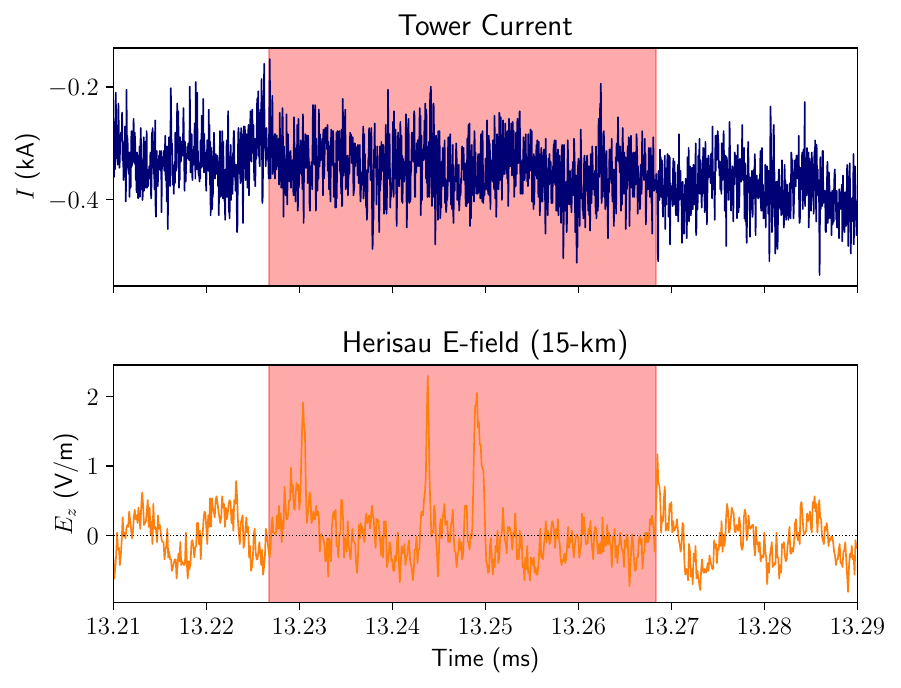}
        \subcaption{Zoom on the four Category B pulses occurring during the frame to the left, including the two shown in Figure~\ref{fig:cbps1}.}\label{fig:cbps2}
    \end{subfigure}
\caption{Category B pulses (4 out of 12) associated with the stepping of a downward-connecting recoil leader prior to the Type-1 Main Pulse ($t_{SL}=13.46$~ms) of Flash \#8. (a) The HSC frame containing the pulses. The yellow arrow indicates propagation direction, and the horizontal black dashed line the approximate cloud base ($\sim$900~m above the Tower). (b) E-field waveforms (bottom panel) of four consecutive Category B pulses. No appreciable current associated with these pulses was measured (top panel). The red shaded region indicates the approximate temporal width of the HSC frame in (a).
}\label{fig:CatBpulses}
\end{figure}


\subsection{Physical Mechanism Comparison}\label{sec:PhysMech}

As discussed in Section~\ref{sec:int}, there has been debate as to whether or not the discharge processes involved in the B and L phases of downward negative leaders are the same \citep{clarence_preliminary_1957, proctor_vhf_1988}.
Despite the clear similarities between Category A/B and Classical/Narrow PBPs, the degree to which the physical mechanisms at play are comparable is less evident.
Nevertheless, some important common ground can be inferred.

\subsubsection{Category A vs. Classical PB Pulses}\label{sec:classicPhysMech}

Aside from the initiation point differences (in-cloud vs. tower tip) mentioned in Section~\ref{sec:companal}, the initial symmetry between upward and downward negative leaders readily explains the observed similarities between the pulses associated with their initial breakdown and propagation.
These further suggest that the difference in conditions, specifically the availability of free charges in upward lightning due to the presence of a conducting ground and tower, as opposed to initiation within the thundercloud for downward lightning, may not play as important a role as one might expect, at least during the Breakdown phase.
Note that the absence of the ``initial E-change (IEC)'' observed by \cite{marshall_electromagnetic_2014} is to be expected here, as the IEC dampens with distance, such that it is effectively flat at 15~km.

\subsubsection{Category B vs. Narrow PB Pulses}\label{sec:narrowPhysMech}

Since all observed E-field pulses are necessarily generated by an acceleration of charge, the lack of any corresponding current pulses measured at the tower base in Category B electric field pulses suggests that they are due to electrical activity disconnected from the stepping of the upward negative leader responsible for the generation of Category A pulses.
Analysis of the HSC frames available for Flash \#8 supports this interpretation. 
Figure~\ref{fig:CatBpulses} shows that the late stage Category B pulses starting at about 13.23 ms are generated  by the stepping of a downward-connecting recoil leader prior to the main pulse. 
(See \cite{oregelchaumont_underlying_2025} for further discussion of how the Type-1 Main Pulse of this particular UPF appears to be triggered by the downward-stepping recoil leader connecting to the main current-carrying channel.)\footnote{Note that Type 2 UPF Category B pulses are likely generated by a similar process that \textit{doesn't} connect to the main channel (or at least not in the same way).} 
To the best of our knowledge, this is the clearest optical-electrical confirmation of a connection between the two phenomena to date, and is consistent with the observed domination of Category B and Narrow PB pulses at later times, when recoil/downward-connecting leaders become more common.

In addition to their resembling Category B pulses in duration and timing during the flash (late in leader development, preceding large current pulses), the aforementioned association of Narrow PBPs with the stepping of a downward negative leader approaching the ground in a downward negative flash is physically analogous to the stepping of the downward-connecting recoil leader in UPF \#8, especially considering the altitude at which both phenomena occur.
In short, given (i) the similarities between Category B and narrow PB pulses, and (ii) the Category B association with recoil leader connection, it is reasonable to conclude that Narrow PBPs could also be generated by a recoil leader in a downward CG flash.

It is important to remember, however, that only one flash recorded in this study had optical observations, and that the camera's $41~\mu$s exposure time, though relatively short, is still long enough that the 12 late-time Category B pulses were often grouped into single frames ($\sim 2-3$ pulses per frame), as shown in Figure~\ref{fig:CatBpulses}.
It is for this reason that the approximate 2D step length of $l_S \gtrsim 56 \pm 4$ m was estimated by dividing the recoil leader's measured channel length by the number of associated pulses, as described in Table~\ref{tab:pulseStats}.
Furthermore, this is only one possible source of Category B pulses: those occurring at earlier times, especially when overlapping Category A pulses, are much more likely associated with the UNL stepping process, like narrow PBPs optically-confirmed to be generated by the stepping of downward negative leaders, as discussed in Section~\ref{sec:companal}.

\subsection{Validation of field--current Relationship Models for Classical PBPs}\label{sec:valid}

\begin{figure}[h]
\centering
\includegraphics[width=\textwidth]{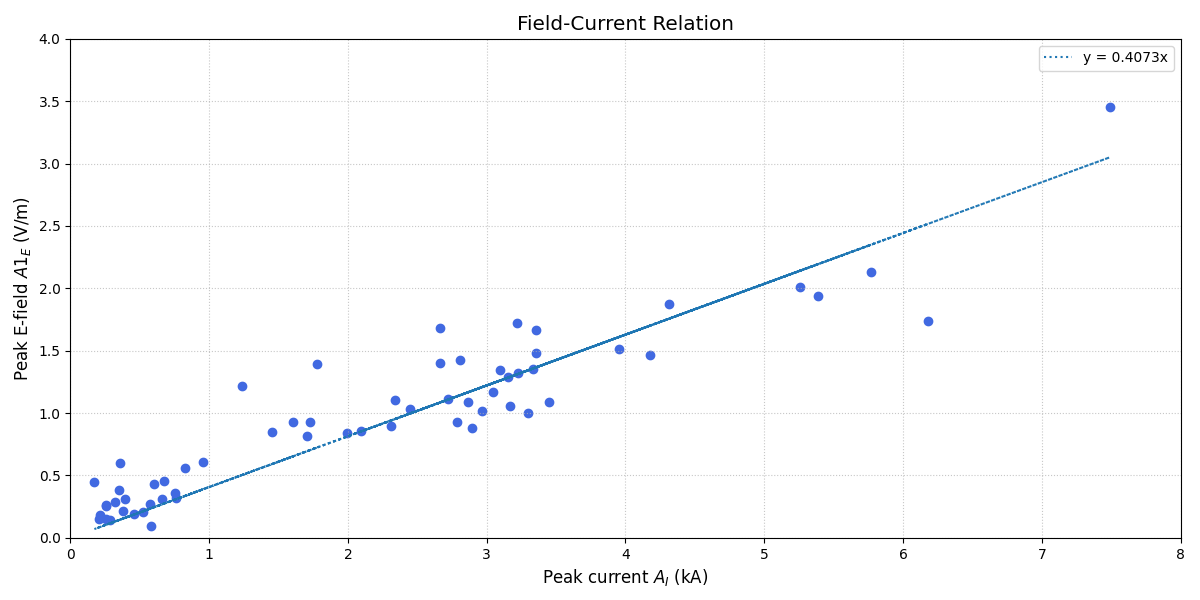}
\caption{Peak electric field versus peak current for bipolar Category A pulses. }\label{fig:EF-Cur}
\end{figure}

The simultaneous measurements of current and electric field waveforms provide a unique opportunity to validate existing models of the field--current relationship for PBPs. \cite{kaspar_model_2017} proposed a physical model describing the PBP stage in downward negative lightning based on a predefined thundercloud charge structure, and deduced a proportionality between the peak currents and electric field amplitudes.
As presented in Figure~\ref{fig:EF-Cur}, we found a very strong linear correlation ($\rho = 0.93$, $r_s = 0.91$, $\tau = 0.76$) between peak E-field and current amplitudes, with a best-fit $R^2 = 0.95$. 


After accounting for the known E-field enhancement factor of $k=1.8$ due to the topography between Mt. S\"antis and Herisau \citep{li_lightning_2016}, our measured peak E-field and current data of Category A pulses yielded $A/I_P \approx 0.23\pm0.05$~V/m/kA, which is consistent with the field--current ratio of $A/I_\mathrm{PBP} \approx 0.163\pm0.025$~V/m/kA predicted by the model proposed in \cite{kaspar_model_2017}. 
Our validation of this relationship using simultaneous measurements lends further support to the hypothesis of analogous physical processes for upward and downward negative leaders.

This field--current correlation also confirms suspicions of a different physical mechanism for Category B pulses, as their normalised E-field amplitudes of $\sim 0.22+$~V/m would imply corresponding current pulse amplitudes ranging from 440 A \citep{kolmasova_subionospheric_2016} to 1.35 kA \citep{kaspar_model_2017}, depending on which field--current relation is used, which we do not observe (see Figure~\ref{fig:cbps2}).






\section{Summary and conclusions}\label{sec:sumcon}




This study investigated stepped leader pulses associated with upward positive flashes at the S\"antis tower and revealed strong similarities with preliminary breakdown pulses (PBPs) observed in downward negative flashes. The key findings can be summarised as follows:

\begin{enumerate}
    \item The waveform characteristics and temporal behavior of Category A pulses associated with upward negative stepped leaders are very similar to ``Classical'' PBPs in downward lightning, supporting the interpretation that the latter are generated during the initial breakdown stage associated with downward negative leader formation;
    \item Category B pulses, whose lack of current correlation suggests an association with disconnected processes, were revealed by high-speed camera imaging in one case to be the result of a downward-stepping recoil leader, which, in combination with their similarity to ``Narrow'' PBPs in downward lightning, indicates that at least a subset of the latter may also be produced by recoil leader connections;
    \item Significant correlations were observed among various bipolar Category A pulse parameters, including linear relationships between (a) first and second E-field half-cycle durations ($R^2 = 0.89$), (b) half-cycle peak amplitudes ($R^2 = 0.90$), (c) E-field peak and maximum derivative (also observed in Category B pulses), and (d) peak E-field and current amplitudes ($R^2 = 0.95$);
    \item These simultaneous measurements of currents and E-fields associated with Category A pulses in upward negative leaders were found to be consistent with the peak field--current relationship proposed by \cite{kaspar_model_2017}, providing quantitative support for this theoretical model of PBPs;
    \item The temporal evolution of pulse types--initial coexistence of Categories A and B followed by a predominance of Category B--mirrors a pattern observed in downward flashes, supporting the bimodal stepping model proposed by \cite{petersen_high-speed_2015}.
\end{enumerate}

\noindent These conclusions should be viewed in light of several limitations.
The number of flashes with complete current, field, and high-speed video records is small, and the association of individual pulses with specific luminous features is constrained by camera spatio-temporal resolution.
It is also important to acknowledge that in-cloud PBPs in downward flashes occur under different geometry, altitude, and ambient conditions than those near the tip of a grounded tower, which may influence breakdown thresholds, streamer inception, and leader stepping dynamics.
Although the present findings are not meant to be interpreted as evidence of identical discharge physics under all atmospheric conditions, they nevertheless have important implications for lightning research: 
despite physical differences in initiation environments (e.g., the presence of free charges in upward lightning from the conducting ground and tower, versus initiation within the thundercloud for downward lightning), the demonstrated similarities between upward and downward negative leaders 
suggest common underlying physical mechanisms, especially in the case of high-altitude initiated upward flashes, where the breakdown voltage is similar, as in the case of the S\"antis tower.

The availability of simultaneous channel-base current and remote electric field measurements, supplemented by high-speed imaging, makes tower-initiated upward flashes a potential proxy for investigating aspects of downward flash initiation that are otherwise inaccessible to conventional field measurements alone.
Such measurements provide valuable constraints on still-debated models of lightning initiation and early leader development.
Future studies incorporating larger and more complete datasets (particularly for Category B pulses), VHF interferometric observations, and/or 3D reconstructions are recommended to explore the distinct physical origins of these pulse types.
Coupled with refined modelling of leader stepping and field–current relationships, these efforts will serve to further refine our understanding of the mechanisms underlying these processes and their implications for lightning initiation theory.



\backmatter


\bmhead{Acknowledgements}
The authors would like to thank Florent Aviolat for developing a data-visualisation software that expedited the identification of events, Kevin Abou Jaoude for his assistance with pulse analysis, and Hannes Kohlmann for the providing the EUCLID/ALDIS data for prior activity verification.

\section*{Declarations}

\subsection*{Funding}
This work was supported in part by the Swiss National Science Foundation (Project no.s 200020\_175594 and 200020\_204235) and the European Union's Horizon 2020 research and innovation program (grant agreement no. 737033-LLR).

\subsection*{Competing interests}
None.
\subsection*{Data availability}
All processed data analysed during this study are included in this published article. 
Raw data sets generated during the current study are available from the corresponding author on reasonable request. 
\subsection*{Code availability}
Upon request from corresponding author. 

\subsection*{Authors' contributions}
T.O.C. wrote the manuscript and completed the data analysis and processing begun by M.A., who also gathered the data alongside M.R. and F.R., heads of the S\"antis Lightning Research Facility, during the summer 2014 experimental campaign. A.S. gathered the data during the summer 2021 campaign (alongside M.R. and F.R.) and designed the diagrams. All authors reviewed the manuscript.

\bibliography{CatABvPBPs}

\end{document}